\documentclass[aps,nofootinbib,prd,eqsecnum,showpacs,showkeys,preprintnumbers]{revtex4-1}
\usepackage[caption=false]{subfig}
\usepackage{graphicx}
\usepackage{amsmath}
\usepackage{amsfonts}
\usepackage{amssymb}
\usepackage{color}
\usepackage{bm}
\usepackage{mathrsfs}
\usepackage{epstopdf}
\usepackage{url}
\usepackage{footnote}
\usepackage{textcomp}
\usepackage{ulem}
\usepackage{esint}
\usepackage[unicode=true, pdfusetitle,
bookmarks=true,bookmarksnumbered=false,bookmarksopen=false,
 breaklinks=false,pdfborder={0 0 1},backref=false,colorlinks=false]{hyperref}
\usepackage{multirow}
\usepackage{pifont}

\begin{document}

\title{Cosmological constraints on dynamical dark energy model in $F(Q)$ gravity}

\author{O. Enkhili}
\email{omarenkhili@gmail.com}
\author{S. Dahmani}
\email{dahmani.safae.1026@gmail.com}
\author{D. Mhamdi}
\email{dalale.mhamdi@ump.ac.ma}
\author{T. Ouali}
\email{ t.ouali@ump.ac.ma}
\author{A. Errahmani}
\email{ ahmederrahmani1@yahoo.fr}
\date{\today }
\affiliation{Laboratory of Physics of Matter and Radiation, Mohammed I University, BP 717, Oujda, Morocco}

\begin{abstract}
Extended teleparallel gravity, characterized by $F(Q)$ function where $Q$ is the  non-metricity scalar, is one of the most promising approaches to general relativity. In this paper, we reexamine a specific dynamical dark energy model,  which is indistinguishable from the $\Lambda$CDM model at present time and exhibits a special event in the future, within $F(Q)$ gravity. To constrain the free parameters of the model, we perform a Markov Chain Monte Carlo (MCMC) analysis, using the last data from Pantheon$^{+}$ and   the latest measurements of the H(z) parameter  combined. On the basis of this analysis, we have find that our dynamical dark energy model, in the context of F(Q)  gravity, lies in the quintessence regime rather than in the phantom regime as in the case of general relativity.  Furthermore, this behaviour affects the future expansion of the Universe as it becomes decelerating at $1\sigma$ confidence level for $z<-0.5$ and showing a bounce at $z_{\text{B}}\approx -0.835$. Finally, we have support our conclusion with a cosmographic analysis.\\
\end{abstract}


\maketitle
\tableofcontents
\section{introduction}
The study of the accelerated expansion of the Universe in the late-time is one of the most interesting areas of modern cosmology. This phenomenon has been confirmed and supported by different cosmological observations, namely the Supernovae Type Ia (SNIa) measurements \cite{perlmutter1999measurements,riess1998observational}, cosmic microwave background \cite{CMB1, CMB2}, baryon acoustic oscillations \cite{BAO1,BAO2} and large scale structure \cite{LS1, LS2}. To explain the acceleration phase of the Universe, an exotic negative-pressure component, dubbed dark energy (DE) is added to the Universe's energy budget in the context of general relativity. Dark energy causes repulsive gravity behavior on large cosmological scales, initially described by the cosmological constant, $\Lambda$ \cite{einstein1917kosmologische} with an equation of state (EoS) parameter, $\omega=-1$. This model, called Lambda Cold Dark Matter ($\Lambda$CDM), is the  most accepted by recent observational data \cite{1A, 2A, 3A}. Unfortunately, the $\Lambda$CDM  model has some serious theoretical problems, including fine-tuning \cite{RevModPhys.61.1, Peebles_2003, fin} and coincidence problem \cite{con1, con2} in addition to various observation issues, called cosmological tensions \cite{T1, T2, T3}.
Several dynamical dark energy models have been proposed in order to overcome these problems, such as the quintessence \cite{martin2008quintessence,chiba2000kinetically}, k-essence \cite{armendariz2001essentials,chimento2004power,malquarti2003new}, Chaplygin Gas \cite{kamenshchik2001alternative,zhang2006interacting}, holographic dark energy \cite{bouhmadi2018more,li2004model,belkacemi2020interacting,belkacemi2012holographic,bargach2021dynamical}, generalized holographic dark energy \cite{granda2008infrared,bouhmadi2011cosmology,belkacemi2012holographic,bouhmadi2018more}  and phantom dark energy where the EoS parameter is slightly less than $-1$ \cite{ph,dahmani2023constraining,R1}. Phantom DE models have some drawback features as their energy density increases with the scale factor and becomes infinite at some point in the future and leads the Universe to end in a Big Rip singularity \cite{ph}. Among the models proposed to cure this drawback, we cite the Little Sibling of the Big Rip (LSBR), which has been proposed as one of the mildest models of the possible apocalypse and describes perfectly the current acceleration of the Universe \cite{bouhmadi2015little}. Several works have studied this model in $F(R)$ cosmology \cite{LSBRH} and in quantum cosmology ~\cite{LSBR5, LSBR8} (see also \cite{LSBR3, LSBR4, LSBR6, LSBR7}). Moreover, it has recently been shown that this model is well supported by cosmological observations data \cite{amine}, and can reduce the Hubble tension compared to the $\Lambda$CDM model \cite{safae}. \\ 
Another way of explaining this late acceleration is to modify the structure of the Einstein-Hilbert gravitational Lagrangian itself by considering scalar tensor theories \cite{Amendola1999,errahmani2006high} or by replacing the Ricci scalar with a single-variable $f(R)$ \cite{SV}, two-variable $f(R,T)$ \cite{TTV,Errahmani:2024ran} or even three-variable \cite{TV}  $f(R,T,P)$, where $R$ is the Ricci scalar,  $T$ is the trace of the energy-momentum tensor and $P$ is the Kretschmann scalar.  These approaches have led to many important extensions of general relativity, including $F(R)$ gravity \cite{starobinsky1987new}, $F(G)$ gravity \cite{de2009construction}, where G is the Gauss–Bonnet term, $F(P)$ gravity \cite{erices2019cosmology}, Horndeski scalar-tensor theories \cite{deffayet2009covariant} etc. However, from a general differential geometric point of view, and taking into account the affine properties of  manifolds, the curvature is not the only geometric object that may be used within a geometrical framework to construct gravitational theories. Indeed, torsion and non-metricity are two other  essential geometric objects connected to a metric space, along with the curvature. They can be used to obtain $F(\mathcal{T})$ \cite{bamba2013effective} and  $F(Q)$ \cite{nester1998symmetric, jimenez2018coincident,mandal2023cosmological} gravity theories, where $\mathcal{T}$ and $Q$ are the torsion and the non-metricity scalar, respectively.\\

 Recently, many different applications of $F(Q)$ gravity  have been studied. Among these, the authors of  \cite{lazkoz2019observational}, have found an interesting result  for the current value of the Hubble parameter, which was close to the Planck's estimate, by considering the $F(Q)$ Lagrangian as a polynomial function of redshift.  To demonstrate the late-time acceleration of the Universe, a new parametrization of the EoS parameter in the context of $F(Q)$, was established in \cite{koussour2023observational}.
Thanks to the energy conditions, the authors of \cite{mandal2020energy}  have restricted the families of  $F(Q)$ models compatible with the accelerated expansion of the Universe. A new class of $F(Q)$ theories, in which the non-metricity scalar is coupled non-minimally to matter Lagrangian  has been proposed by \cite{harko2018coupling}, where the cosmological solutions  are considered in two general classes of models, and are characterized by the accelerated expansion at late time. In addition, the authors of \cite{MhamdiEPJC2024}  have shown that the $F(Q)$ model is well supported by the background and perturbation data, using a particular form of $F(Q)$. For more studies related to  the applications of $F(Q)$ gravity  see for instance  \cite{mandal2023cosmological,cosmology,anagnostopoulos2021first,anagnostopoulos2023new,mandal2022reply,
khyllep2021cosmological,beh2022geodesic,MhamdiEPJC2024}.\\

In this paper, we aim to reexamine the current accelerated expansion of the Universe by considering the specific dynamical dark energy (DDE) model \cite{bouhmadi2015little}, in the context of $F(Q)$ gravity.   Indeed, in the context of general relativity, this kind of dark energy mimics a phantom like behaviour and smooth the big rip singularity in a good agreement with the observational data \cite{dahmani2023constraining,LSBR3,amine,safae}. Furthermore, This dark energy model predicts more dark matter in the past \cite{safae} and reproduces $\Lambda$CDM in the current time. As we will show, some of these properties will undergo a change in the context of $F(Q)$ gravity \cite{R2,R3,R4}. Among these: this dark energy will behave as a quintessence, the expansion will decelerates and the Universe will faces a bounce in the future. \\

In this line, we estimate  constraints of the cosmological parameters of our setup. We first perform cosmological analysis with Markov Chain Monte Carlo (MCMC) approach \cite{MCMC}, using the most recent data from Supernovae type Ia, namely Pantheon$^{+}$ \cite{Pan} combined with $H(z)$ data \cite{7}. We then perform a statistical comparison between our model and  $\Lambda$CDM by using the Akaike Information Criterion (AIC) \cite{akaike1974new,AIC} and the Bayesian Information Criterion (BIC) \cite{BIC}. In addition, we present a cosmographic analysis of the model, including the evolution of the deceleration parameter (q), jerk parameter (j), and snap (s) parameter \cite{1, 2, 3}.\\

The outline of this article is as follows: In Section \ref{S1}, we give a brief review of the $F(Q)$ gravity, in Section \ref{S2}, we present the modified Friedman equations in the $F(Q)$ gravity. In Section \ref{S3}, we describe the methodology of the datasets. In Section \ref{S4} we present the results and discussions. Section \ref{S5} is devoted to the cosmographic analysis of the model.  Finally, in Section \ref{S6} we summarize our results.

\section{Overview of ${F}(Q)$ gravity}\label{S1}
In this section, we give a brief introduction to the formalism of ${F}(Q)$ gravity. We consider ${F}(Q)$ modified gravity, in which the basic object is the non-metricity tensor $Q_{\lambda\mu\nu}$, given by \cite{nester1998symmetric}
\begin{equation}
\Delta_{\lambda}g_{\mu\nu}=Q_{\lambda\mu\nu}.
\label{EQ1}
\end{equation}
The non-metricity scalar, which is an important quantity in this theory, is defined in terms of the disformation tensor, $L_{\mu\nu}^{\lambda}$, as follows
\begin{equation}
Q=-g^{\mu\nu}(L_{\beta\nu}^{\alpha} L_{\nu\alpha}^{\beta}-L_{\beta\alpha}^{\alpha}L_{\mu\nu}^{\beta}),
\label{EQ11}
\end{equation}
where  the disformation tensor is symmetrical with respect to lower indices and is defined by
\begin{equation}
L_{\mu\nu}^{\lambda}=-\frac{1}{2}g^{\lambda\gamma}(Q_{\mu\gamma\nu}+Q_{\nu\gamma\mu}-Q_{\gamma\mu\nu}).
\end{equation}
In terms of the non-metricity conjugate, the non-metricity scalar is given by
\begin{equation}
Q=-P^{\alpha\beta\gamma}Q_{\alpha\beta\gamma},
\end{equation}
where the non-metricity conjugate, $P^{\alpha\beta\gamma}$, is defined in terms of the non-metricity tensor and its two independent traces, $Q_{\alpha}={{Q_{\alpha}}^{ \beta}}_{\beta}$ and $\tilde{Q}^{\alpha}={Q_{\beta}}^{\ \alpha\beta}$,
as 
\begin{equation}
    P^{\alpha}_{\mu\nu}=\frac{1}{4}\left(-Q^{\alpha}_{\mu\nu}+2 Q^{\alpha}_{(\mu\nu)}-Q^{\alpha}g_{\mu\nu}-\tilde{Q}^{\alpha}g_{\mu\nu}-\delta^{\alpha}_{(\mu} Q_{\nu)}\right).
\end{equation}
The action of the $F(Q)$ modified gravity is written as follows \cite{jimenez2018coincident}
\begin{equation}
S=\int {d^4x\sqrt{-g}}\left(-\frac{1}{2\kappa^2}{F}(Q)+\mathcal L_m \right ),
\end{equation}
where ${F}(Q)$ is a function {of} the non-metricity scalar $Q$, g is the determinant of the metric $g_{\mu\nu}$ and $\mathcal L_m$ is the Lagrangian density of matter.
The field equation of the $F(Q)$ gravity is obtained by varying the action with respect to $g_{\mu\nu}$, and it takes the following form
\begin{equation}
-\frac{2}{\sqrt{-g}}\nabla_{\alpha}(\sqrt{-g}F_QP^{\alpha}_{\mu\nu} )
 +\frac{1}{2}g_{\mu\nu}F+F_Q(P^{\alpha\beta}_{\nu}Q_{\mu\alpha\beta}-P^{\alpha\beta}_{\mu}Q_{\alpha\beta\nu})=\kappa^2 T_{\mu\nu}, 
\end{equation}
where $F_Q = \frac{\partial F}{\partial Q}$, and the energy-momentum tensor, $T_{\mu\nu}$, is given by
\begin{equation}
    T_{\mu\nu}=-\frac{2}{\sqrt{-g}}\frac{\delta\sqrt{-g}\mathcal L_m}{\delta\sqrt{g_{\mu\nu}}}.
\end{equation}
By varying the action with respect to the affine connection \cite{mandal2023cosmological}, we obtain the following equation
\begin{equation}
    \nabla_{\mu}\nabla_{\nu}(\sqrt{-g}F_QP^{\mu\nu}_{\alpha})=0.
\end{equation}
\section{cosmological model}\label{S2}
To examine the cosmological consequences of the $F(Q)$ gravity, we consider the Friedmann-Lemaître-Robertson-Walker (FLRW) line element, which describes the flat, homogeneous, and isotropic Universe, is given by
\begin{equation}
ds^{2}=-dt^{2}+a^{2}(t)\left(dx^{2} +dy^{2}+dz^{2}\right ),
\end{equation}
where $t$ is the cosmic time, $x$, $y$ and $z$ denote the Cartesian coordinates, $a(t)$ is the cosmic scale factor, and the Hubble parameter $H(t)$ is defined by $H(t) = \frac{\dot a}{a}$, with $\dot{a}$  denotes the derivative of the scale factor with respect to the cosmic time. 
\subsection{The generalized Friedmann equations}
In the FLRW geometry,  we get the non-metricity scalar as $Q = 6H^2$. We consider the matter content of the Universe as consisting of a perfect and isotropic fluid, with the energy momentum tensor given by
\begin{equation}
T_{\mu\nu}=(p+\rho)u_{\mu}u_{\nu}+pg_{\mu\nu},
\end{equation}
where $p$ and $\rho$ are the pressure and the energy density of the fluid, respectively, $u_{\mu}$ is the four velocity vector normalized according to
$u^{\mu}u_{\nu}=-1$.\\

To better fulfill our goal, we  consider the splitting of ${F} (Q)$ as  ${F (Q) =Q + f(Q)}$.  Using the FLRW metric, we get two Friedmann equations as \cite{jimenez2018coincident,cosmology}
\begin{eqnarray}
 -\frac{f}{2}+3H^{2}+Qf_Q&=&\rho,    \label{21}\\ 
 (12f_{QQ}H^{2}+1+f_{Q})\dot H&=&-\frac{\rho+p}{2},
\label{22}
\end{eqnarray}
where we have set $\kappa^2=1$. According to the above equations, we recover the standard model, for $f=0$ (i.e. $F(Q)=Q$). The same equations, Eqs. (\ref{21}) and (\ref{22}) can be rewritten as
\begin{equation}
    \left \{
\begin{array}{c} 
3H^{2}=\rho+\rho_{DE},\label{5}\\
-2 \dot H-3H^{2}=p+p_{DE},
\end{array}
 \right.
\end{equation}
where we have introduced a geometrical dark energy density, $\rho_{DE}$, and its pressure, $p_{DE}$, as follows
\begin{equation}
    \left \{
\begin{array}{c} 
\rho_{DE}=\frac{f}{2}-Qf_Q,\label{6}\\
p_{DE}=2 \dot H(f_Q+2Qf_{QQ})-\frac{f}{2}+Qf_Q.
\end{array}
 \right. 
\end{equation}

Furthermore,   the effective equation of state  using Eq. (\ref{5}) can be written as,
\begin{equation}
\omega_{\text{eff}}=\frac{-2 \dot H-3H^{2}}{3H^{2}}.
\end{equation}

\subsection{Dynamical dark energy}

In order to understand the characteristic properties of the dark energy driving the recent acceleration of the Universe,  we need to   parameterize its equation of state (EoS). To this aim, we consider one of the main studied dynamical dark energy  given by \cite{bouhmadi2015little}
\begin{equation}
    \rho_{\mathrm{DE}}+p_{\mathrm{DE}}=-\frac{A}{3},\label{7}
\end{equation}
where  $\rho_{\mathrm{DE}}$, $p_{\mathrm{DE}}$, and $A$ are the energy density,  the pressure, and a constant characterizing the dark energy model, respectively.\\

The primary motivation for taking into account this parametrization form is its advantage to smooth the big rip singularity. Indeed, in this scenario, the size of the observable Universe and its expansion rate grow to infinity while the cosmic time derivative of the Hubble parameter converges to a constant value. Furthermore, this model deviates slowly from the standard model, by the constant $A$ and it behaves as phantom like dark energy models. The equation of state of this dynamical dark energy model, dubbed Little Sibling of the Big Rip (LSBR) in \cite{bouhmadi2015little},  is given by
\begin{equation}
\omega_{\mathrm{DE}}=-(1+\frac{A}{3(\rho_{{\mathrm{DE}},0}+A\ln{(\frac{a}{a_0})})}),
\label{d1}
\end{equation}
according to Eq. (\ref{d1}), the DDE model describes a phantom DE, for positive values of $A$ (i.e. $\omega_{\mathrm{DE}}<-1$). For negative values of $A$, the model describes a quintessence DE (i.e. $\omega_{\mathrm{DE}}>-1$) and mimics the standard model in the limit $A\rightarrow 0$. Despite the aforementioned advantages, The DDE model faces some drawbacks. Indeed,  This dynamical dark energy model is unable to describe the  very early expansion of the Universe as its energy density becomes negative i.e. it is not supported by high redshifts \cite{safae}.\\

Using Eqs. (\ref{6}) and (\ref{7}), we get
\begin{equation}
    2 \dot H(f_Q+2Qf_{QQ})=-\frac{A}{3}, \label{8}
\end{equation}
the dot (.) indicates the derivative with respect to cosmic time.   Using the relation between the  derivative of $H$  and  the non-metricity scalar $Q$ 
\begin{equation}
    \dot H=H \frac{dH}{dx}=\frac{1}{12}\frac{dQ}{dx}, \label{9}
\end{equation}
where $x=\ln{(a)}$, Eq. \eqref{8} becomes 
\begin{equation}
    \frac{1}{6}\frac{dQ}{dx}(f_Q+2Qf_{QQ})=-\frac{A}{3}. \label{10}
\end{equation}
In order to illustrate our purpose, we consider the following form of $f(Q)$ \cite{mandal2023cosmological}
\begin{equation}
    f(Q)=6 \gamma H_0^2(\frac{Q}{Q_0})^{n},\label{11} 
\end{equation} 
where $\gamma$, $n$ and $Q_0=6H_0^2$ are  constants. 
In this case, Eq. \eqref{10} becomes
\begin{equation}
    \frac{{n}}{2}\frac{dQ}{dx}(\frac{Q}{Q_0})^{n-1}\gamma (2n-1)=-A,\label{14}
\end{equation}
we also can write this equation in the form
\begin{equation}
    \frac{d(\frac{Q}{Q_0})^{n}}{dx}=-\frac{2A}{Q_0\gamma (2n-1)}.
\end{equation}
By integration, we obtain
\begin{equation}
  (\frac{Q}{Q_0})^{n}=-\frac{2A}{Q_0\gamma (2n-1)}x+1.  \label{16}
\end{equation}
Combining Eqs \eqref{11} and \eqref{6} gives the form of $\rho_{\mathrm{DE}}$ in terms of $Q$ as
\begin{equation}
   \rho_{\mathrm{DE}}= \gamma Q_0 (\frac{1}{2}-n)(\frac{Q}{Q_0})^{n},\label{17}
\end{equation}
which in terms of $x$, it becomes
\begin{equation}
\rho_{\mathrm{DE}}= Ax+\gamma Q_0 (\frac{1}{2}-n) \label{rhoDE}.
\end{equation}
We finally obtain the Friedmann equation, from Eq. \eqref{5}, as follows
\begin{equation}
    E^2=\Omega_{m_0}(1+z)^3+\Omega_{r_0}(1+z)^4-{\Omega_{x}}\ln{(1+z)} +{\Omega_{\Lambda}} \label{FriedmannEq}
\end{equation}
where  $E = H/H_{0}$ is the dimensionless Hubble rate, $\Omega_x=2A/Q_0$ and $\Omega_\Lambda=\gamma (1-2n)$ and we have used $x=-\ln(1+z)$. \\
The last two terms in Eq. (\ref{FriedmannEq}) represent the dimensionless parameters of dark energy density, which can be expressed as
\begin{equation}
    \Omega_{DE}=-\Omega_{x}\ln(1+z) +\Omega_{\Lambda} \label{OmegaDE}
\end{equation}

From Eq. \eqref{FriedmannEq}, the model predicts an increase in dark matter density as \( z \rightarrow \infty \). However, in the future, the model is dominated by dark energy, and its behavior is characterized by the sign of \(\Omega_{x}\). Eq. (\ref{FriedmannEq}) can be approximated as follows: $E^2 \approx -\Omega_{x}\ln(1+z)+\Omega_{\Lambda}$. For $\Omega_{x}$ being positive, the Hubble rate \( H \) diverges while its derivative \( \dot{H} \) remains constant \cite{bouhmadi2015little}. This abrupt event, termed  "Little Sibling of the Big Rip", smooths the big rip singularity in the future. This phenomenon has been extensively studied in references \cite{safae,bouhmadi2015little,LSBR5,LSBR8}. However, if $\Omega _{x}$ is negative, the Universe's dynamical behavior undergoes a bounce, as we will illustrate in section \ref{S4}. From now on, we will refer to our setup as $F(Q)$-DDE.

\section{Statistical analysis}\label{S3}
In order to obtain optimal constraints on cosmological parameters of the $F(Q)$-DDE model and compare them with those of the $\Lambda$CDM model, we perform a Markov Chain Monte Carlo (MCMC) analysis, using the last data from Pantheon$^+$ and Hubble parameter data (H(z)). We consider a parameter space composed of the Hubble constant, $H_0$, the density parameter of total matter, $\Omega_m$, the two parameters that describe the $F(Q)$ gravity, ,$\gamma$ and $n$, as well as the DDE parameter, $A$, by means of the dimensionless parameter, $\Omega_x=2A/Q_0$.  The prior related to the free parameters is given in Tab.  (\ref{prior}) 

\begin{table}[!htp]
\centering
\begin{tabular}{c|c}
\hline
\multicolumn{1}{c|}{\bf Parameters} & \multicolumn{1}{c}{\bf Prior}\\
\hline      
$\Omega_{\textrm{m}}$   & [$0, 1$]
 \\[0.1cm]
$H_0$ &[$40, 100$]
 \\[0.1cm]
 
$n$ &[$-1,1$]
 \\[0.1cm]
$\gamma$ &[$-1,1$]
 \\[0.1cm]
$\Omega_{\text{x}}$ &[$-1,1$]
 \\[0.1cm]

$M_B$ & [$-20,-19$]
 \\[0.1cm]
\hline
\end{tabular}
{\caption{Prior imposed on different parameters for the $F(Q)$-DDE and $\Lambda$CDM models.}\label{prior}}
\end{table}

\subsection{Pantheon$^+$ dataset}
We use the latest Pantheon$^+$ measurements \cite{Pan}, composed of 1701 light curves of 1550 distinct Supernovae Type Ia (SNIa) confirmed by spectroscopy, ranging in $z$ interval, $z\in[0.001, 2.26]$.\\
The $\chi_{\text{Pan}^+}$ function is written as
\begin{equation}
\chi_{\text{Pan}^+}^2=\Delta\mu^T\mathcal{M}_{\text{Pan}^+}^{-1}\Delta\mu,
\end{equation}
where $\mathcal{M}_{\text{Pan}^+}$ is the covariance matrix and $\Delta\mu$ is a vector where each of its elements refers to the $i^{\text{th}}$ distance modulus of SNIa,  given by
\begin{equation}
\Delta\mu^i=
\begin{cases}
\mu^i-\mu^i_{\text{ceph}}, &   i\in \text{ Cepheid host},\\
\mu^i-\mu_{\text{theo}}(z_i), &    \text{ Otherwise},
\end{cases}
\end{equation}
where $\mu^i_{\text{ceph}}$ is the distance modulus for the Cepheid calibrated host-galaxy of the $i^{th}$ SNIa provided by SH0ES \cite{SH0ES} and  $\mu_{\text{theo}}$ is the theoretical distance modulus, defined as\\
\begin{equation}
\mu_{\text{theo}}(z_i)=m_b-M=5\log_{10}{\frac{d_L}{\text{Mpc}}}+25,
\end{equation}
where $M$ is the absolute magnitude, $m_b$ is the apparent magnitude and $d_L$ is the luminosity distance, defined as
\begin{equation}
d_L=(1+z) c\int_{0}^{z}\frac{dz}{H(z)},
\end{equation}
 $c$ is the speed of light. 
\subsection{H(z) Data}

Secondly, we use Hubble expansion rate data \cite{7} to impose tighter constraints on our F(Q)-DDE  model. Generally, $H(z)$ data can be obtained either through the clustering of galaxies and quasars, measured via Baryon Acoustic Oscillations (BAO) in the radial direction \cite{data2}, or through the differential age method \cite{data3}. In our current analysis, we have employed a compilation of 36 data points of the Hubble parameter, which exhibit no correlation. The $\chi^2_{{\text{H(z)}}}$ function is given by
\begin{equation}
\chi^2_{{\text{H(z)}}}=\sum_{i=1}^{36}\left[\frac{H_{\text{theo}}(z_i)-H_{\text{obs}}(z_i)}{%
\sigma_i}\right]^2,
\end{equation}
where $H_{\text{theo}}(z_i)$ is the theoretical values, $H_{\text{obs}}(z_i)$ is the observed values at $z_i$ and $\sigma_i$ is the standard deviation.\\

To analyze both the H(z) data and type Ia supernovae samples simultaneously, we use the total Chi-square function  $\chi^2_{\text{tot}}$ given by
\begin{equation}
\chi^2_{\text{tot}}=\chi^2_{\text{Pan+}}+\chi^2_{\text{H(z)}}.
\end{equation}

\subsection{Information criteria}
The aim of this work, besides the parameter estimations, is to select the best model  supported by the observation data. In fact, a model with a smaller value of $\chi^2$ indicates that the model is well fitted to the observation data. However, a large number of parameters can increase the quality of the fit and therefore lead to a smaller $\chi^2$. In this case, the $\chi^2$ statistics is not an appropriate way to compare models. To confidently predict the goodness of the fit between the $F(Q)$-DDE and $\Lambda$CDM models, we use another statistical tool, namely the Akaike Information Criterion (AIC) \cite{akaike1974new,AIC}, which depends on the number of parameters, $\mathcal{N}$,  given by
\begin{equation}
AIC = \chi^2_{min}+2\mathcal{N}.
\end{equation}
As $\chi^2$ statistics, a model with a smaller value of AIC indicates that the model is most supported by  the observation data. For this we calculate\footnote{In our analysis, we have considered the $\Lambda$CDM  model as a reference model.}
\begin{equation}
\Delta AIC =  AIC_{\text{model}}- AIC_{\Lambda CDM}.
\end{equation}
The model selection rule of $\Delta$AIC, is as follows: for 0$\leqslant\Delta$AIC$<$2, it suggests that the model has nearly the same level of support from the observation data as $\Lambda$CDM, for 2$\leqslant\Delta$AIC$<$4, it indicates the model has slightly less support compared to the $\Lambda$CDM model, for  4$\leqslant\Delta$AIC$<$10  the data still support the  model but less than the preferred one and for  $\Delta AIC>10$,  the model is not supported by the observational data compared to the $\Lambda$CDM model.\\

We also calculate the  Bayesian information criterion (BIC), given by \cite{BIC}
\begin{equation}
BIC = \chi^2_{min}+\mathcal{N}\log{(k)},
\end{equation}
 where a smaller value of BIC indicates that the model is most supported by the observation data. For this we calculate
\begin{equation}
\Delta BIC =  BIC_{\text{model}}- BIC_{\Lambda CDM}.
\end{equation}
A positive (negative) value of $\Delta$BIC shows that the $F(Q)$-DDE ($\Lambda$CDM) model is preferred. The selection rules of BIC is as follows: for  0$\leqslant\Delta$BIC$<$2   means that the  evidence is insufficient, for 2$\leqslant\Delta$BIC$<$6, indicates  that there is a positive evidence while for $6\leqslant\Delta$BIC$<10$ indicates a solid evidence.

\section{Results and discussions}\label{S4}
\begin{figure}[htbp]
\centering
\includegraphics[width=0.7\textwidth]{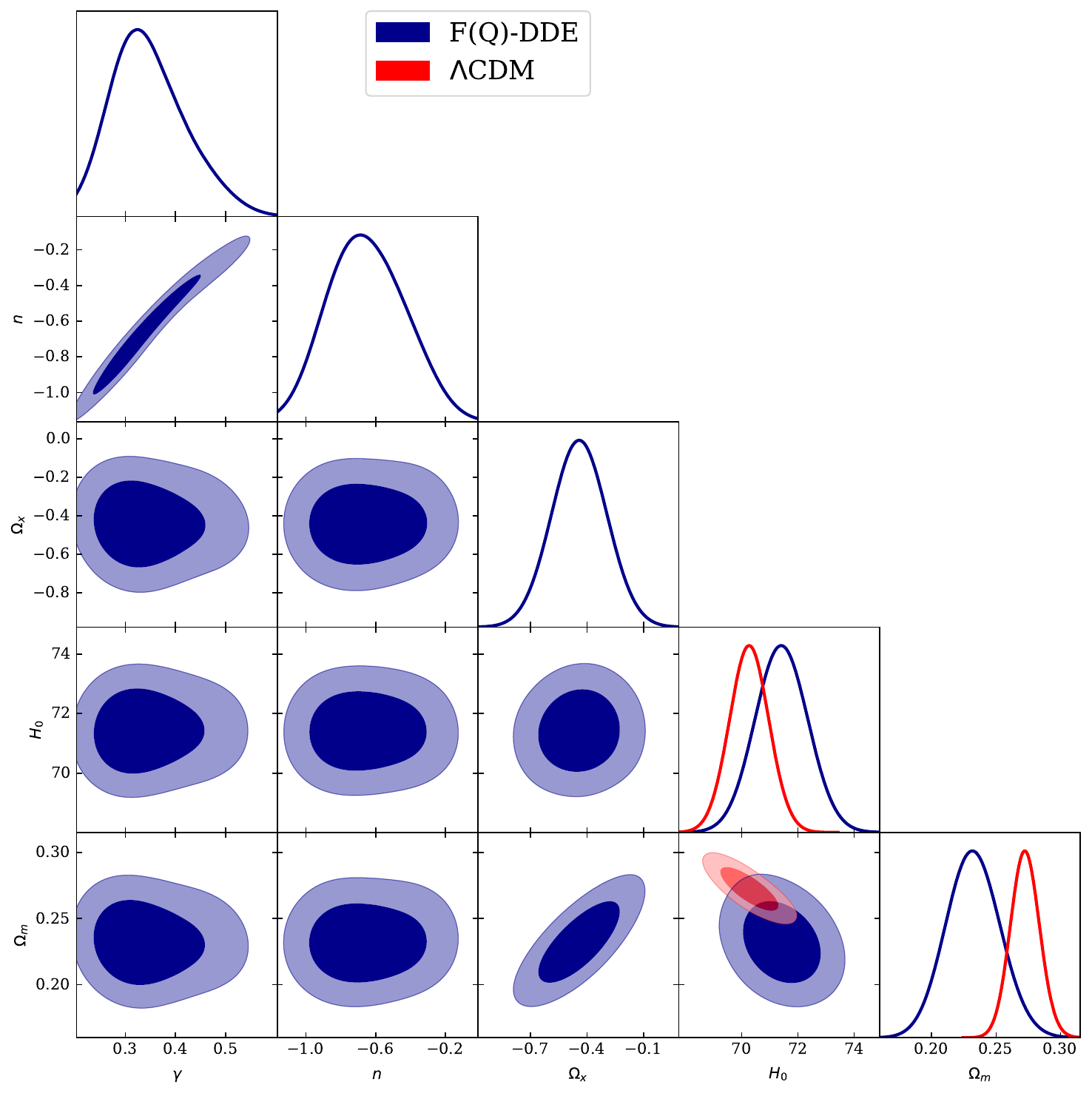}
\caption{The 1$\sigma$ and 2$\sigma$ confidence contours and the 1D posterior distributions obtained for  the F(Q)-DDE and $\Lambda$CDM models, using Pantheon$^+$ and H(z) combined. The figure is obtained with the Getdist package \cite{Get}.}
\label{CF1} 
\end{figure}

In Table (\ref{parameters}), we present the 68\% C.L. constraints on the cosmological parameters of the $\Lambda$CDM and $F(Q)$-DDE models obtained using combination of Pantheon$^+$ and H(z) data. Fig. (\ref{CF1}) shows the 1D posteriori distributions and 2D marginalized contours at $1\sigma$ and $2\sigma$ for the $\Lambda$CDM and $F(Q)$-DDE models. From Table (\ref{parameters}), the Hubble constant values are $H_0=71.42\pm 0.88$ km/s/Mpc for the $F(Q)$-DDE model and $H_0=70.27\pm 0.66$ km/s/Mpc for $\Lambda$CDM at 68\% C.L.. The latter is smaller, at around $1.04\sigma$  from the value obtained by $F(Q)$-DDE. In Table (\ref{parameters}), we also list constraints on the two parameters describing the $F(Q)$ gravity, i.e. ($\gamma$, $n$)=( $0.346^{+0.061}_{-0.087}$,  $-0.65^{+0.21}_{-0.24}$). We observe a strong correlation between these two parameters  in  Fig. (\ref{CF1}). Furthermore, we notice that in the $F(Q)$-DDE gravity, the DDE parameter  value is negative i.e.  $\Omega_{\text{x}}=-0.44\pm 0.14$ at 68\% C.L., which from  Eq. (\ref{d1}) leads to $\omega_{\text{de}}>-1$. This shows that, in the context of $F(Q)$ gravity, the DDE model  behaves like a quintessence one, contrary to the results found in \cite{bouhmadi2015little,LSBR3,amine,safae} where it is shown that  observational data prefer the phantom like behaviours of DDE.\\
Moreover, the future evolution of the Universe depends strongly on the sign of $\Omega{x}$, $\gamma$ and $n$. Indeed, in the future (i.e. $-1<z<0$), the asymptotic behavior of the Friedmann equation (\ref{FriedmannEq}) can be simplified to  
\begin{equation}
    E^2\approx-\Omega_{x} \ln{(1+z)} +\gamma (1-2n),\label{AsFriedmannEq}
\end{equation}
and $E^2$ is an increasing function of $z$. This means that $E^2$ decreases as the Universe expands and, consequently, the Universe will bounce (i.e. $E^2 = 0$) at some point in the future where the redshift can be approximated to $\ln(1+z_{B})\approx \gamma (1-2n)/\Omega_{x}$. Using the numerical values, from Table (\ref{parameters}), for $\Omega_{x}$, $\gamma$ and $n$, the redshift $z_{B}$   corresponding to the bounce can be estimated to $z_{B}\approx -0.835$. This scenario can be interpreted as follow: the  dark energy density, described by Eq. \ref{rhoDE} and which drives the current accelerated expansion of the Universe, decreases as the Universe expands.  At this point, the Universe undergoes a bounce and starts contracting. While in the context of general relativity, this dark energy density smooths the big rip singularity.\\

In the same table, we present the value of $\chi_{\text{min}}^2$, $\Delta$AIC as well as $\Delta$BIC. According to the $\chi_{\text{min}}^2$ value, we see that the DDE model in the context of $F(Q)$ gravity increases the goodness of fit to the observational data over $\Lambda$CDM, with $\Delta\chi_{\text{min}}^2=\chi_{\text{model}}^2-\chi_{\Lambda\text{CDM}}^2=-10.76$. However, a large number of parameters can lead to a smaller value of $\chi^2$ and thus increase the quality of the fit. To confidently predict the goodness of fit between models, we use the Akaike Information Criterion (AIC) and Bayesian information criterion (BIC), which depend on the number of parameters. According to Tab. \ref{parameters}, we obtain $\Delta\text{AIC}=\text{AIC}_{\text{model}}-\text{AIC}_{\Lambda\text{CDM}}=-6.76$, which gives a statistical preference for our model. However, the positive sign of $\Delta\text{BIC}$ shows a preference for the $\Lambda$CDM model over our model  as this criterion penalizes models with additional parameters.\\

\begin{figure}[htbp]
\centering
\includegraphics[width=0.48\textwidth]{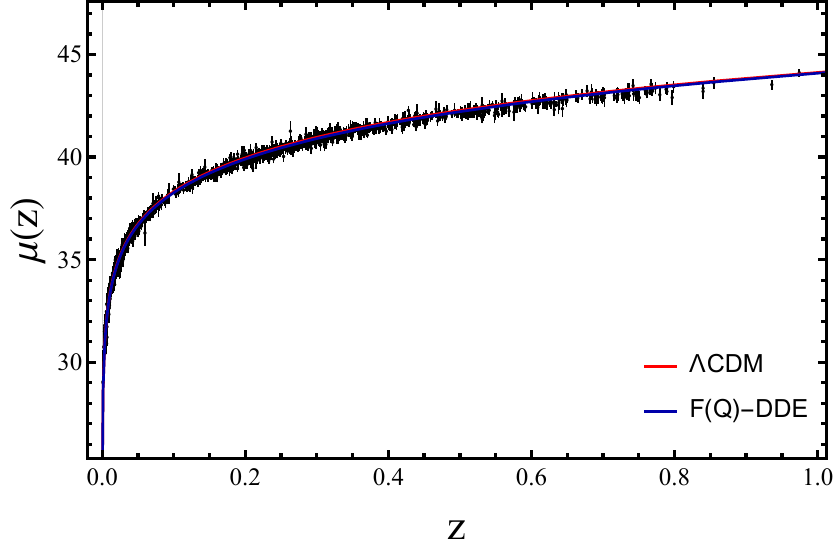}
\includegraphics[width=0.48\textwidth]{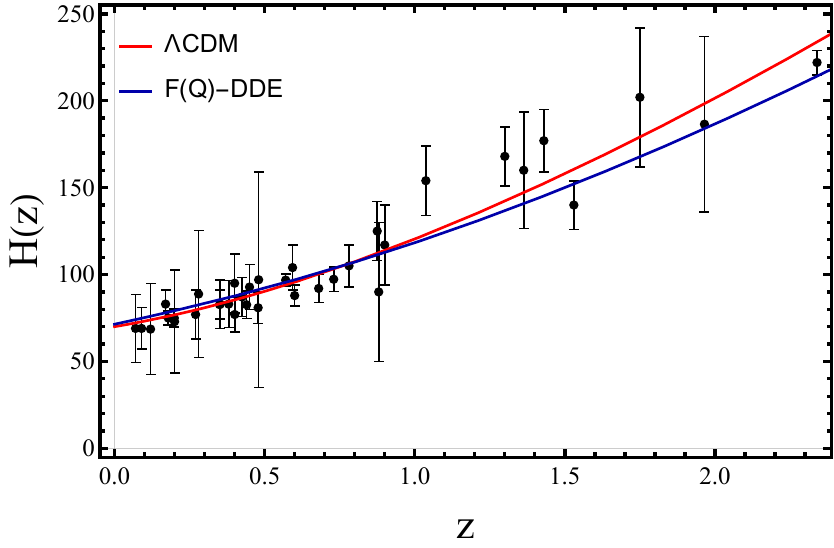}
\caption{The figure shows the error bar of 1701 data points from the Pantheon$^{+}$, together with the fit of the $\mu(z)$ function with respect to redshift  for $F(Q)$-DDE and $\Lambda$CDM (left panel) and the error bar of 57 points from the H(z) datasets, including the fit of Hubble function $H(z)$ versus redshift  for $F(Q)$-DDE and $\Lambda$CDM  (right panel).}
 \label{mu} 
\end{figure}

Fig. (\ref{mu}) shows the theoretical evolution of the distance modulus $\mu(z)$ (left panel)  and the Hubble function $H(z)$ (right panel), predicted by the F(Q)-DDE model as a function of $z$, using the results presented in Table (\ref{parameters}). For comparison, both data samples are also presented in the same figure according to the theoretical prediction of the $\Lambda$CDM model. This figure shows that the F(Q)-DDE model behaves identically to  $\Lambda$CDM, and that both models are in good agreement with the observational data. 
\begin{table}[htbp]
\centering
\begin{tabular}{c|c|c}
\hline
\hline
\textbf{Parameters} & \textbf{$\Lambda$CDM} & \textbf{$F(Q)$-DDE} \\ 
\hline
$\Omega_m$         &   $ 0.281\pm 0.011$            &  $0.232\pm 0.020$\\ 
$H_0$              &         $70.27\pm 0.66$         &   $71.42\pm 0.88$ \\ 
$\gamma$           &  - &   $ 0.346^{+0.061}_{-0.087}$\\ 
$n$                &  - &          $-0.65^{+0.21}_{-0.24}$ \\ 
$\Omega_{\text{x}}$                & -  &          $-0.44\pm 0.14$ \\ 
$M$                &$-19.322 \pm 0.023$ & $-19.306\pm 0.025$\\ 
\hline
\multicolumn{3}{c}{\bf  statistical results}\\
\hline
$\chi_{\text{min}}^2$           & $1564.11$ & $            1553.35$ \\ 
AIC          & $1570.11$ & $1563.35$ \\ 
$\Delta$AIC          & $0$ & $-6.76$ \\ 
BIC         & $1586.48$ & $1590.64$ \\ 
$\Delta$BIC          & $0$ & $+4.16$ \\ 
\hline
\hline
\end{tabular}
\caption{Summary of the mean$\pm 1\sigma$ values of the cosmological parameters for the $\Lambda$CDM and $F(Q)$-DDE models, using  Pantheon$^+$ and H(z) combined, as well as the $\chi_{\text{min}}^2$, $\Delta$AIC and $\Delta$BIC.}
\label{parameters}
\end{table}

\section{Cosmographic analysis }\label{S5}
\subsection{Cosmographic parameters}
\begin{figure}[htbp]
\centering
\includegraphics[width=0.49\textwidth]{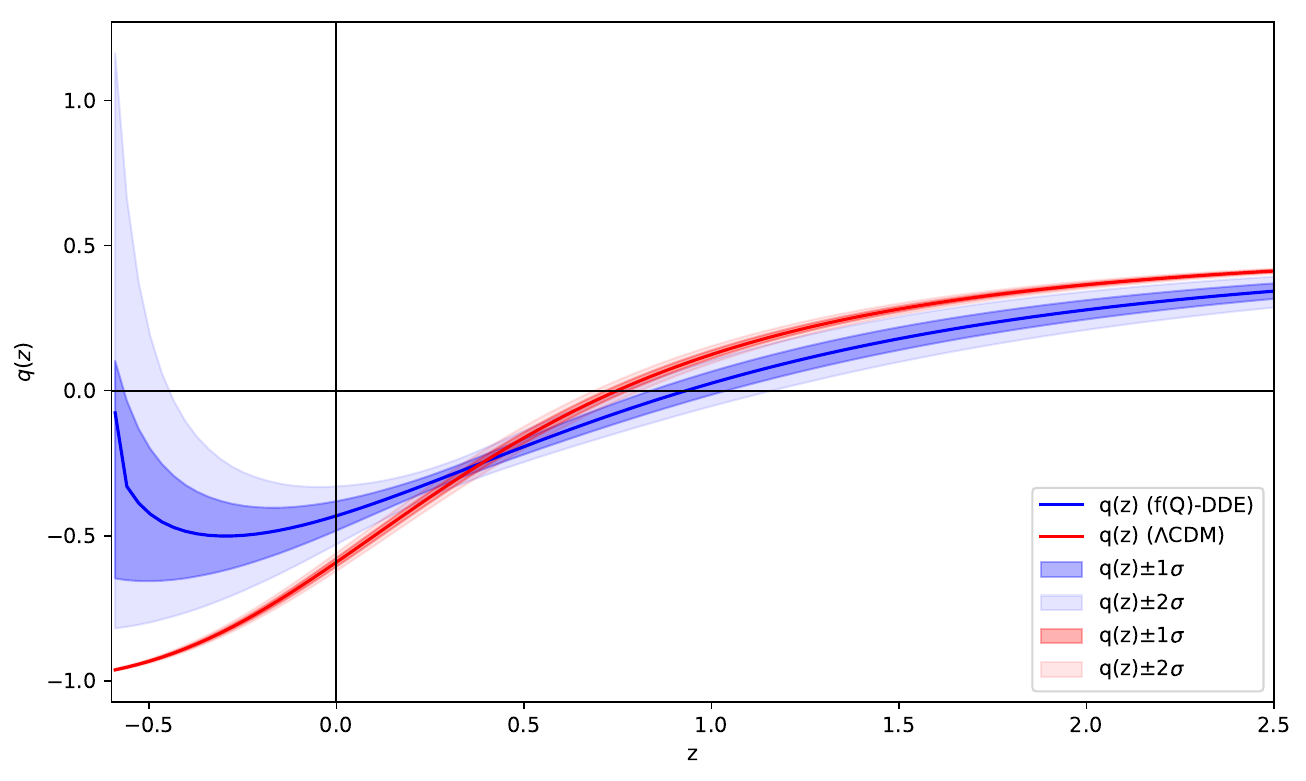}
\includegraphics[width=0.49\textwidth]{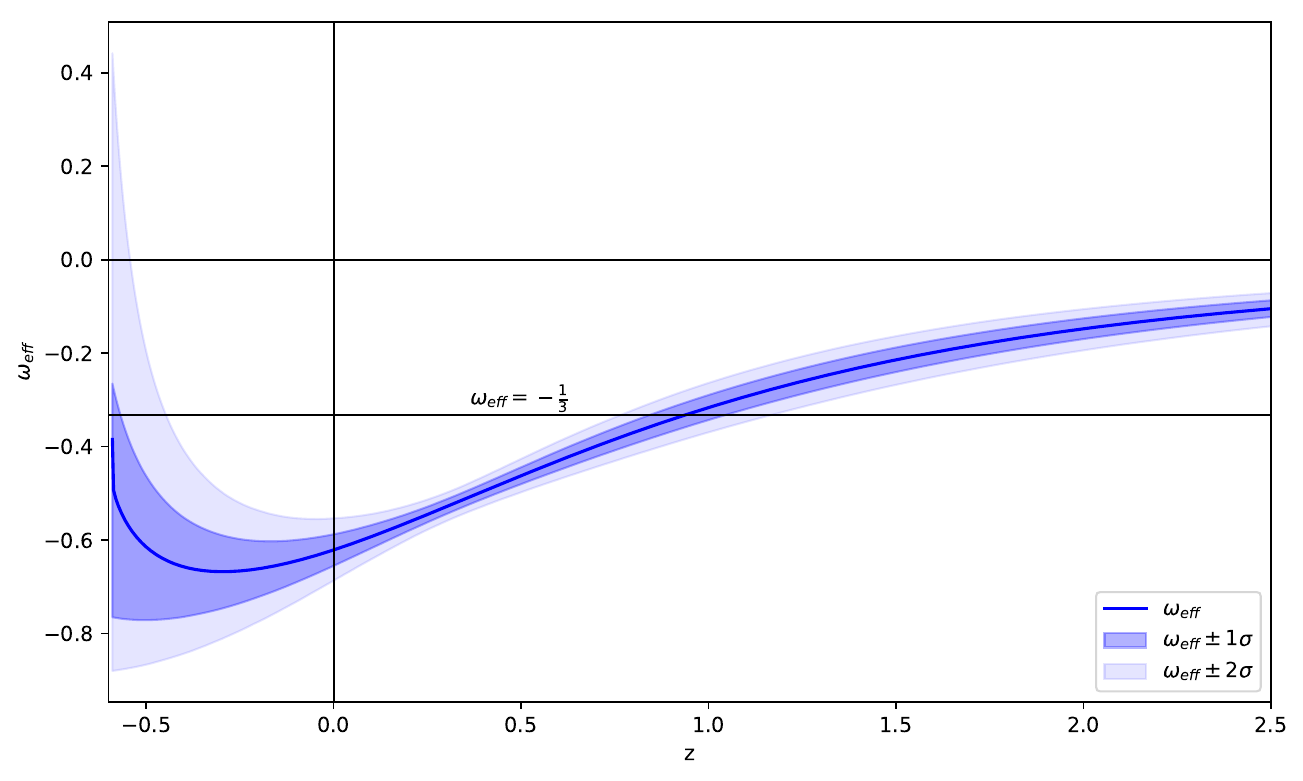}
\caption{Evolution of q(z) in a confidence interval $1\sigma$ and $2\sigma$ for the $\Lambda$CDM and $F(Q)-DDE$ models (left panel) and the evolution of $\omega_{\text{eff}}$ for the  F(Q)-DDE model, using the chains of the free parameters obtained by Pantheon$^+$ and H(z) combined (right panel).}
 \label{q} 
\end{figure}

Given the large number of cosmological models that have been proposed to explain the late accelerated expansion of the Universe, 
a mathematical tool, without assuming a dark energy models, is useful to discriminate between all proposed dark energy models. An approach called cosmography has been proposed for this reason \cite{COS, COS1, COS2}. This analysis makes it possible to compare several cosmological models using the derivatives of scale factor $a$ by means of the Taylor series expansion of the scale factor \cite{3, H1, H2}.\\
The aim of this part is to study the Taylor series expansion of the scale factor $a$ with respect to cosmic time, $t$, by introducing the following cosmographic parameters: the Hubble parameter, deceleration parameter, jerk parameter and snap parameter, defined respectively as follows\\
\begin{equation}
H(t)=\frac{1}{a}\frac{da}{dt},
\end{equation}
\begin{equation}
q(t)=-\frac{1}{a{H^{2}}}\frac{d^2a}{dt^2},
\end{equation}
\begin{equation}
j(t)=\frac{1}{a{H^{3}}}\frac{d^3a}{dt^3},
\end{equation}
\begin{equation}
s(t)=\frac{1}{a{H^{4}}}\frac{d^4a}{dt^4}.
\end{equation}
 Fig. \ref{q}, shows the evolution of the mean value of the deceleration parameter $q(z)$ and its uncertainty at 1$\sigma$ and 2$\sigma$ as a function of redshift for the $F(Q)$-DDE and $\Lambda$CDM models (left panel) and the evolution of $\omega_{\text{eff}}$ for the  $F(Q)$-DDE model (right panel). The $q(z)$ parameter measures the rate at which the expansion of the Universe is accelerating or decelerating. In particular, positive values of $q(z)$ indicates that the expansion is decelerating, which also corresponds to $\omega_{\text{eff}}>-\frac{1}{3}$, while negative values of $q(z)$ indicate that the expansion is accelerating and $\omega_{\text{eff}}<-\frac{1}{3}$. Fig. (\ref{q}) shows that the expansion of the Universe is characterized by a transition phase, from deceleration to acceleration in the recent past. For our model, the phase transition (i.e. $q(z_{\text{tr}})=0$) occurs at redshift $z_{\text{tr}}=0.938$, while the redshift transition for  $\Lambda$CDM  is less than that of $F(Q)$-DDE, i.e. at $z_{\text{tr}}^{\Lambda\text{CDM}}=0.728$. For our model, we obtain the actual value of the deceleration parameter, $q_0=-0.4315\pm 0.0507$, with a difference of $3.01\sigma$ compared to the $\Lambda$CDM model, where $q_0=-0.5914\pm 0.0158$.  The most important result  in Fig. (\ref{q}) is that a phase transition  in the future from accelerating to decelerating expansion at the $1\sigma$ confidence level, in the redshift  $z<-0.5$, is possible. This result is also confirmed by the equation of state, $\omega_{\text{eff}}$, where $\omega_{\text{eff}}>-\frac{1}{3}$ at $1\sigma$  in the redshift $z<-0.5$ (see the right panel of Fig. (\ref{q})). This conclusion is recently found in \cite{futur}. We also note that  $\omega_{\text{eff}}>-1$ and that the model is in the quintessence region for all redshifts. \\ 
 
\begin{figure}[htbp]
  \centering
  \includegraphics[width=0.48\textwidth]{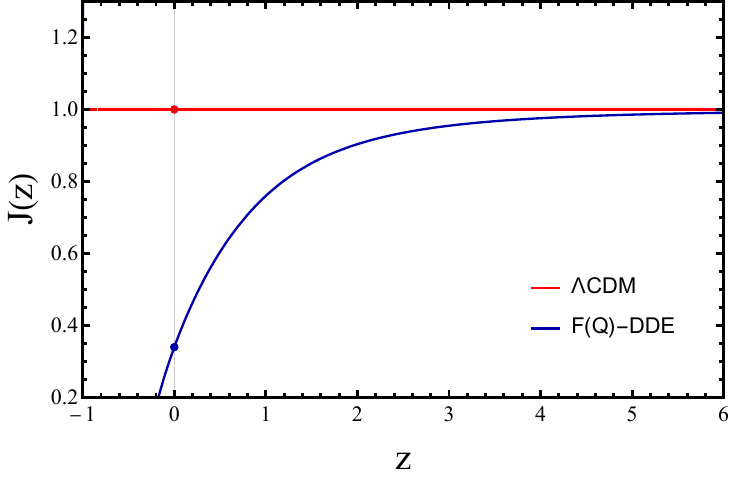}
  \includegraphics[width=0.48\textwidth]{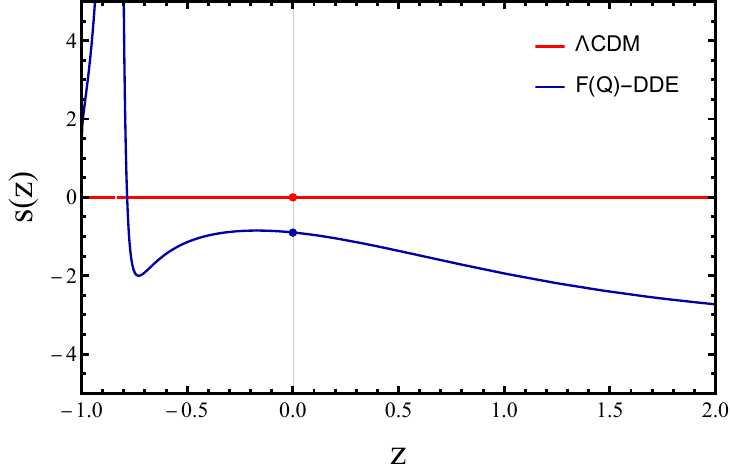}
  \caption{The figure shows  the evolution of the jerk parameter $j(z)$ (left panel) and  the snap parameter $s(z)$ (right panel) as functions of the redshift.}
  \label{jerk} 
\end{figure}

Fig. \ref{jerk}, shows the evolution of the jerk parameter (left panel) and the snap parameter (right panel) as functions of the redshift. The red and blue points represent the present time ($z=0$), for the $\Lambda$CDM and F(Q)-DEE models, respectively. The left panel of Fig. \ref{jerk} shows that the jerk parameter approaches unity, $j\simeq 1$ in the near past (i.e. $z>4$). Therefore, our model behaves identically to the $\Lambda$CDM, as the jerk parameter always remains constant, $j_{\Lambda\text{CDM}}=1$.    At the present time, we obtain a positive value of the jerk parameter $j_0 = 0.3401 \pm 0.2073$ with a difference of  $3.1\sigma$ compared to the $\Lambda$CDM model and we can see a significant deviation between $F(Q)$-DDE  and $\Lambda$CDM for low and negative redshifts (see the left panel of Fig. \ref{q}). The right panel of Fig. \ref{jerk}  shows the evolution of the snap parameter. From this figure, we notice significant deviations between  $F(Q)$-DDE and $\Lambda$CDM for all redshifts. The present value of the snap parameter for our model is $s_0=-0.9804 \pm 0.2043$ with a difference of $4.79\sigma$ compared to the $\Lambda$CDM model.


\begin{table}[htbp]
\centering
\begin{tabular}{c|c|c}
\hline
\hline
\textbf{Parameters} & \textbf{$\Lambda$CDM} & \textbf{$F(Q)$-DDE} \\ 
\hline
$q_{\text{0}}$           & $-0.5914\pm 0.0158$ & $-0.4315\pm 0.0507$ \\ 
$j_{\text{0}}$           & $1$ & $0.3401\pm 0.2073$ \\ 
$s_{\text{0}}$           & $0$ & $-0.9804\pm 0.2043$ \\ 
\hline
\hline
\end{tabular}
\caption{Summary of the mean$\pm 1\sigma$ values of the current cosmographic parameters as derived parameters.}
\label{parameters}
\end{table}

\subsection{Statefinder diagnostic}
\begin{figure}[htbp]
  \centering
  \includegraphics[width=0.48\textwidth]{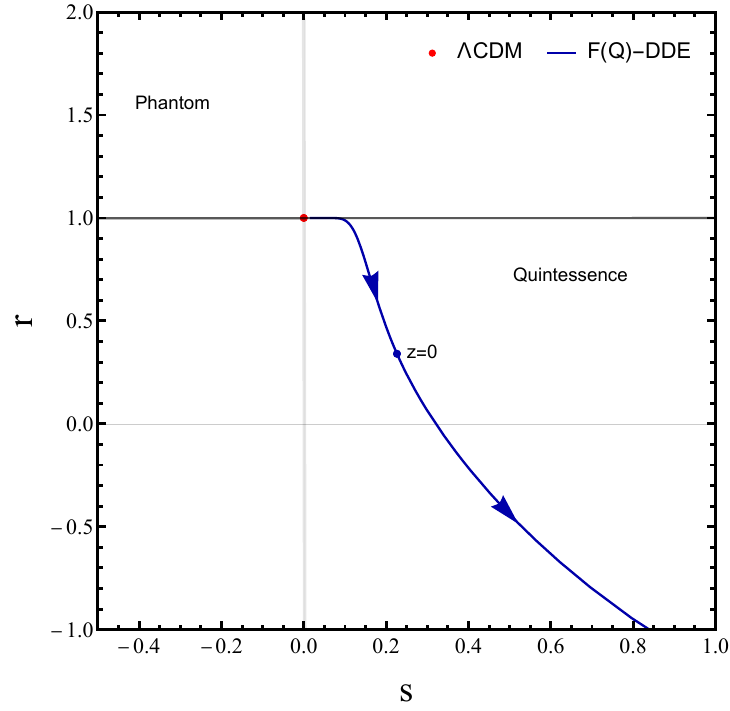}
   \includegraphics[width=0.46\textwidth]{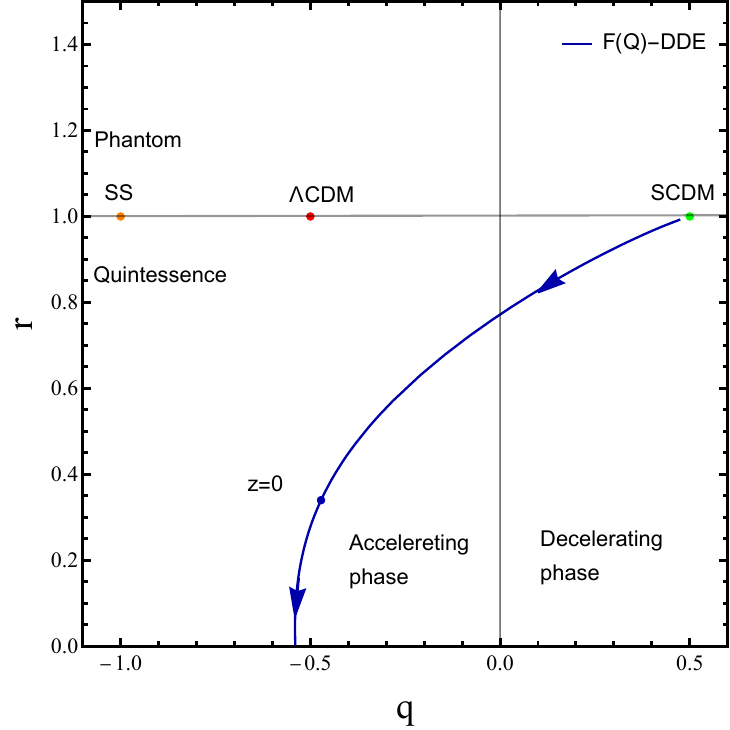}
  \caption{The figure shows the evolutionary trajectory of the  $\left \{ s,r \right \}$ plane (left panel). The statefinder of the $\Lambda$CDM model is indicated by fixed  point ({red color}), and the blue point indicates the $F(Q)$-DDE model in the present time. The right panel presents the evolutionary trajectory of the $\left \{ q,r \right \}$ plane for $F(Q)$-DDE model. }
  \label{rs} 
\end{figure}

In this section, we use statefinder diagnostic technique which is an alternative tools to compare and distinguish between different models of dark energy \cite{ST, ST1}. This geometric technique uses the pair of parameters $\left \{ s,r \right \}$, where $r$ is the jerk parameter and $s$ is snap parameter, defined as follows \cite{ST, ST1}
\begin{equation}
r=\frac{\dddot{a}}{aH^3},
\end{equation}
\begin{equation}
s=-\frac{r-1}{3(q-0.5)}.
\end{equation}
In order to evaluate the behavior of our model, the trajectories of the pairs $\left \{ s,r \right \}$ and $\left \{ q,r \right \}$ are represented in Fig. \ref{rs}, where the arrows represent the time evolution from the past to the future. The left panel of Fig. \ref{rs}  displays the evolutionary trajectory of the  $\left \{ s,r \right \}$ plane. From this figure,    our model is located in the quintessence region (i.e. $r<1$ and $s>0$).  The right panel of Fig \ref{rs} shows the evolutionary trajectory of the $\left \{ q,r \right \}$ plane. We notice that the evolution of our model starts  from the region where the Universe is dominated by matter abbreviated by SCDM ($q = 0.5$ and $r = 1$),  and converges to a point ($q = -0.55$, $r = 0$), representing the late-time evolution of the Universe.

\section{Conclusions}\label{S6}
In this study, we reevaluate the dynamical dark energy (DDE) model in $F(Q)$ gravity. This dark energy model has been previously studied in the context of general relativity, exhibiting phantom behavior \cite{bouhmadi2015little}. The model predicts more dark matter density in the past. However, in the future, the model is dominated by dark energy and exhibits distinctive behavior depending on the sign of  $\Omega_{x}$. Specifically, for a positive value of $\Omega_{x}$, this abrupt event is dubbed "Little Sibling of the Big Rip", as it smooths the big rip singularity. Conversely, this event leads the Universe to bounce when \( \Omega_{x} \) is negative. \\

To understand the nature of the DDE model within $F(Q)$ gravity, we constrained the model parameters by performing a Markov Chain Monte Carlo analysis using a combination of two different background datasets, namely the Pantheon$^+$ and Hubble parameter data. We have found that for the $\Lambda$CDM model, $\Omega_m = 0.281 \pm 0.011$ and $H_0 = 70.27 \pm 0.66$. However, for the $F(Q)$-DDE model, $\Omega_m = 0.232 \pm 0.020$, $H_0 = 71.42 \pm 0.88$, $n = -0.65^{+0.21}_{-0.24}$, and $\gamma = 0.346^{+0.061}_{-0.087}$. In particular, we obtained a negative value for the DDE parameter, $\Omega_{\text{x}} = -0.44 \pm 0.14$ at 68\% C.L., indicating that the DDE model in F(Q) gravity is strictly oriented towards a quintessence region, and in the future the Universe bounce.\\

In addition, we have obtained $\chi_{\text{model}}^2 < \chi_{\Lambda\text{CDM}}^2$, indicating that the model fits the observational data well. 
 However, relying solely on $\chi_{\text{min}}^2$ to compare models with different numbers of free parameters is not enough. Therefore, we also evaluated the AIC and BIC values. Our findings show that $F(Q)$-DDE model is slightly more preferable in terms of AIC compared to $\Lambda$CDM where $\Delta$AIC = -6.76. Conversely, our model is less favored in terms of BIC, where $\Delta$BIC = +4.16 in favor of $\Lambda$CDM. Additionally, the theoretical predictions of the distance modulus $\mu(z)$ and the Hubble function $H(z)$ from the $F(Q)$-DDE model are in good agreement with both data samples.\\

Unlike the case in the context of general relativity, we have noticed that our setup in the context of $F(Q)$ gravity behaves as like a quintessence and faces a bounce in the future at $z_{B}\approx -0.835$. We have found that the phase transition from deceleration to acceleration, in $F(Q)$-DDE, stars late than $\Lambda$CDM. We have also evaluated the current cosmographic parameters including deceleration parameter $q_{0}=-0.4315\pm 0.0507$, jerk parameter $j_{0}=0.3401\pm 0.2073$, and snap parameter $s_{0}=-0.9804\pm 0.2043$. Finally, we have analyzed the statefinder diagnostics in the $\left \{ s,r \right \}$ and $\left \{ q,r \right \}$ planes. We have found that, from statefinder diagnostic $\left \{ s,r \right \}$, our model is located in the quintessence region while from the $\{q,r\}$ plane, the DDE model evolves from a state where the Universe is dominated by matter in the past to a late-time state of the Universe.

\end{document}